\documentclass[aps,pre,twocolumn,superscriptaddress]{revtex4-2}
\usepackage{graphicx}% Include figure files
\usepackage{epsfig}%add
\usepackage{subfigure}  %added
\usepackage{float}
\usepackage{braket}
\usepackage{pslatex,latexsym}
\usepackage{bm,float,lipsum}
\usepackage{amsmath}
\usepackage{amssymb}
\usepackage{amsfonts}
\usepackage{mathrsfs}
\usepackage{multirow}
\usepackage{longtable}
\usepackage{chngpage}
\usepackage{tabulary}
\usepackage{rotating}
\usepackage{color}
\usepackage{rotating}
\usepackage{changes}
\usepackage[colorlinks,
            linkcolor=red,
            anchorcolor=blue,
            citecolor=green
            ]{hyperref}
\begin{document}

\title{Contagion dynamics in time-varying metapopulation networks with node's activity and attractiveness}

\author{Lang Zeng}
\affiliation{School of Physics and Electronic Science, East China Normal University, Shanghai 200241, China}
\affiliation{Department of Mathematics, Nonlinear Dynamics \& Mathematical Application Center, 
	Kyungpook National University, Daegu 41566, Republic of Korea}

\author{Ming Tang}
\thanks{Corresponding author}
\email{tangminghan007@gmail.com}
\affiliation{School of Physics and Electronic Science, East China Normal University, Shanghai 200241, China}
\affiliation{Shanghai Key Laboratory of Multidimensional Processing,East China Normal University, Shanghai 200241, China}

\author{Ying Liu}
\affiliation{School of Computer Science, Southwest Petroleum University, Chengdu 610500, P. R. China}

\author{Seung Yeop Yang}
\affiliation{Department of Mathematics, Nonlinear Dynamics \& Mathematical Application Center, 
	Kyungpook National University, Daegu 41566, Republic of Korea}

\author{Younghae Do}
\thanks{Corresponding author}
\email{yhdo@knu.ac.kr}
\affiliation{Department of Mathematics, Nonlinear Dynamics \& Mathematical Application Center, 
		Kyungpook National University, Daegu 41566, Republic of Korea}

%\author{Ying Liu}
%\affiliation{School of Computer Science, Southwest Petroleum University, Chengdu 610500, P. R. China}

%\author{Jie Zhou}
%\affiliation{School of Physics and Electronic Science, East China Normal University, Shanghai 200241, China}

\begin{abstract}

The metapopulation network model is effectively used to study the spatial spread of epidemics with individuals mobility. Considering the time-varying nature of individual activity and the preferences for attractive destinations in population mobility, this paper develops a time-varying network model in which activity of a population is correlated with its attractiveness. Based on the model, the spreading processes of the SIR disease on different correlated networks are studied, and global migration thresholds are derived. It is observed that increasing the correlation between activity and attractiveness results in a reduced outbreak threshold but suppresses the disease outbreak size and introduces greater heterogeneity in the spatial distribution of infected individuals. We also investigate the impact of non-pharmacological interventions (self-isolation and self-protection) on the spread of epidemics in different correlation networks. The results show that the simultaneous implementation of these measures is more effective in negatively correlated networks than in positively correlated or non-correlated networks, and the prevalence is reduced significantly. In addition, both self-isolation and self-protection strategies increase the migration threshold of the spreading and thus slow the spread of the epidemic. However, the effectiveness of each strategy in reducing the density of infected populations varies depending on different correlated networks. Self-protection is more effective in positively correlated networks, whereas self-isolation is more effective in negatively correlated networks. These findings contribute to a better understanding of epidemic spreading in large-scale time-varying metapopulation networks and provide insights for epidemic prevention and control.

\end{abstract}
\maketitle

\section{Introduction} \label{sec:intro}
Nowadays, people are more mobile than ever, and the availability of transportation options like airplanes and high-speed trains has significantly impacted people's lifestyles and further influenced the spread of epidemics. For example the COVID-19, as it continues to ravage the world today. It began with an outbreak in Wuhan, China, and quickly spreads throughout the country and the world ~\cite{label32,label38,label39}. It is the convenient long-distance migration that breaks the geospatial continuity of epidemic spreading. Therefore, migrations of individuals play an important role in the spread of epidemics. To describe the mobility of individuals, researchers have proposed the metapopulation model (or reaction-diffusion model) ~\cite{label20,label21,label22,label23,label24,label25,label26,label27}. In this model, the entire population is geographically divided into subpopulations, and
the different subpopulations are linked by population migration, thus forming a subpopulation network. Individuals within a subpopulation may interact with each other (e.g., infectious diseases)~\cite{label31}, and these individuals can diffuse to neighboring nodes (subpopulations) under certain rules. This reaction-diffusion process vividly represents spatial propagation of epidemics in the real world.

Many researches of epidemics in complex networks are on static networks, i.e., the connections between nodes do not change over time ~\cite{label20,label21,label40,label41}. However, one of a important feature of realistic networks is that interactions between nodes are not fixed and have a temporal nature. These interactions can be inhibited at certain times and activated at others. Examples include person-to-person email~\cite{label33} and collaborative networks of scientists ~\cite{label34}. Due to the timing of the connected edges, there are many different dynamic behaviors in a time-varying networks. For example, in a static network with three sequentially connected nodes A, B, and C, node A may pass information or disease to node B, which in turn passes it on to node C. However, if the edges are time-varying, such as when the edges between B and C precede the edges between A and C, node C cannot be influenced by node A. With the increasing popularity of Bluetooth, wearable devices, sensors, and other technologies, obtaining timestamp data for building temporal networks has become easier. As a result, researches based on temporal networks have become more enriched~\cite{label1,label2,label3,label4,label5,label6,label7,label9,label10,label17,label18}. Perra et al.~\cite{label17,label18} proposed an activity-driven network model in which connected edges change rapidly over time. The model assigns a specific activity level to each node. The activity level of a node measures the frequency at which individuals in a time-varying network engage in social interactions. They found that the outbreak threshold of an epidemic in a time-varying network is significantly larger than that in a corresponding aggregated network. Since the activity-driven network model has excellent mathematical analysis performance and scalability, it is widely used to analyze dynamic processes in time-varying networks~\cite{label35,label36,label37,label12}.

\begin{figure} [ht!]
	\centering
	\includegraphics[width=\linewidth]{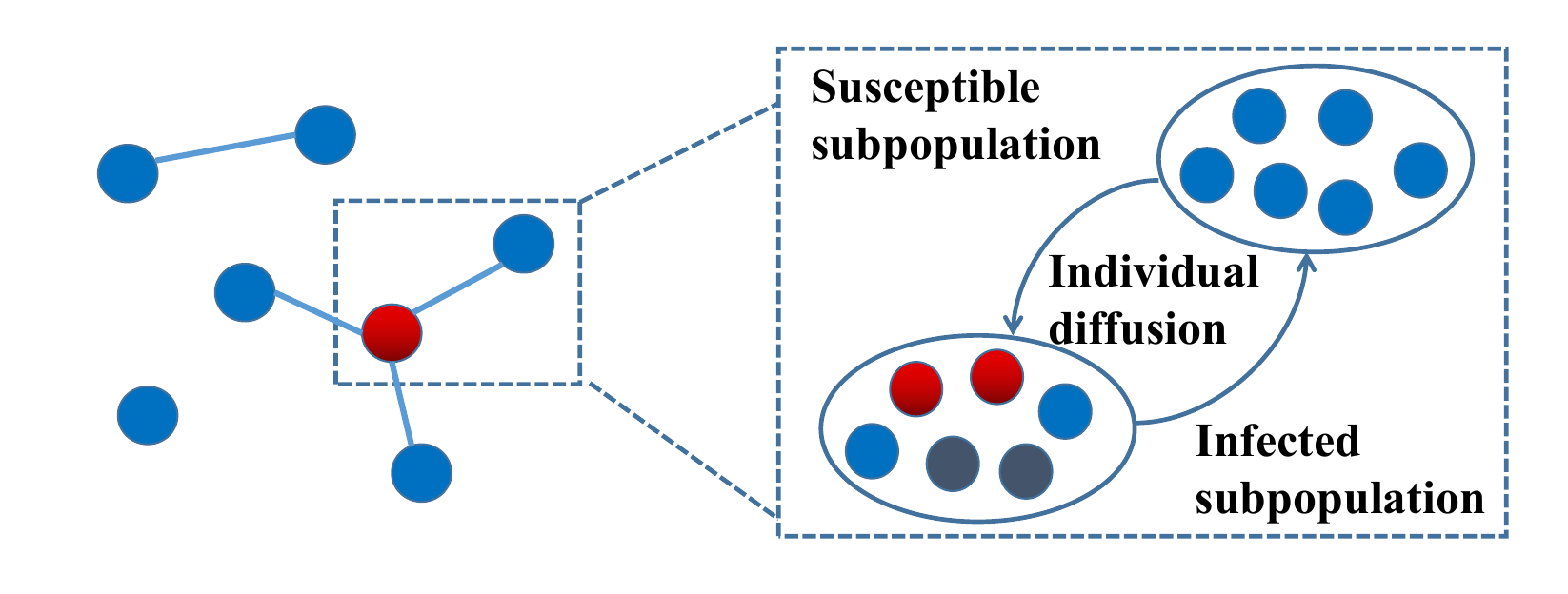}
	\caption{ {\em Schematic diagram of an activity-driven metapopulation model with node attractiveness.} Active nodes generate $l=1$ edge at a time to connect with other nodes. Infected nodes, indicated in red, represent locations experiencing ongoing disease outbreaks, while uninfected nodes, depicted in blue, represent areas without infections.}
	\label{fig}
\end{figure}

In reality, the region of human activity changes during the day and night, and migration paths also change over time. It is necessary to consider the impact of time-varying migration pathways on epidemic spreading, as epidemic characteristics within subpopulations may change with migration.In addition, individuals always travel with a preference or purpose ~\cite{label30}. For example, people prefer to travel to large cities with developed economies (Shanghai and New York) rather than economically undeveloped areas. This suggests
that different locations have different attractions for individuals. People move more frequently to the attractive subpopulations. This type of movement, different from random migration, may lead to different
dynamical behaviors.

Here, we propose a time-varying metapopulation network model where activity of a subpopulation correlates with its attractiveness. The activity of node represents the propensity of a node to establish new connections with others per unit time and the attractiveness of node quantifies the ability of nodes to attract contacts. Each subpopulation $i$ has an activity $a_i$ and an attractiveness $b_i$, which are extracted from the distributions $F(a)$ and $Q(b)$, respectively. The migration paths of individuals are correlated with the activity and attractiveness of the subpopulation. In this paper, we investigate the impact of correlations between activity and attractiveness on the epidemic dynamics on time-varying metapopulation networks. We also explored the impact of non-pharmaceutical interventions on networks with different correlations.
The migration threshold is analytically derived and is validated by numerical simulations. It is found that increasing the correlation between activity and attractiveness leads to lower migration threshold and prevalence, and the spatial distribution of infected individuals becomes more heterogeneous. Implementing
non-pharmacological intervention policies (both selfprotection and self-isolation strategies) in negatively correlated networks is more effective than that in positively correlated networks, and both self-isolation and self-protection increased migration thresholds to slow down the spread of the epidemic. The two strategies performed inconsistently in reducing the density of infected populations, where self-protection is more effective in positively correlated networks and the self-isolation is more effective in negatively correlated networks.
These findings help to understand the impact of correlations
between node attributes on dynamical properties and provide
important insights for epidemic prevention and control.

\section{Time-varying metapopulation network}
\subsection{Network model}
The metapopulation network model is used to study the spread of epidemics due to individual mobility~\cite{label20, label21, label22, label23, label24, label25, label26}. In this model, nodes denote subpopulations (e.g., regions, cities, and countries), and links denote migration routes between subpopulations. Infections occur through the interactions of individuals within subpopulations, and individuals migrate along links between subpopulations, which is called a reaction-diffusion (RD) process~\cite{label20, label21}.

We establish an activity-driven time-varying networks, which is a natural framework for studying dynamical processes that evolve at the same time scale with the networks~\cite{label17,label18}. Consider a metapopulation network with $V$ nodes, the number of particles on each node $i$ is $N_i$, and the total number of particles is $N=\sum_{i=1}^{V} N_i$. Each node is characterized by its activity $a$ and attractiveness $b$,  which represents its propensity to establish links and popularity and are extracted from distributions $F(a)$ and $Q(b)$, respectively. The network generation steps are as follows.
\begin{itemize}
\item[]
(1) At each time step $t$, a metapopulation network $G_t$ is constructed by from $V$ isolated nodes.

(2) Each node $i$ becomes active with probability of $a_{i}\triangle t$ and connects to  $l$ other nodes (both active and inactive
nodes). It chooses node $j$ as the connection target with probability $\frac{b_j}{\sum_{j=1}^{N}b_j}$.

(3) At $t+\triangle t$ all connected edges in the network $G_t$ are removed.
\end{itemize}
In the network $G_t$,  the particles jump out of their nodes at rate $p$, and randomly diffuse to the neighboring nodes along the edges (as illustrated in Fig. \ref{fig}).

The correlation between activity and attractiveness is adjusted in the following way. First, we construct a fully positively correlated network by sorting activity and attractiveness of nodes in ascending order and matching them one by one to assign to all nodes. We use $\triangle_i$  to denote the difference of the order of activity and attractiveness of the $i$th node , and $\triangle_i=0$ at this time. Then, a fraction of $q$ nodes are randomly selected, and their activity and attractiveness are randomly switched to reduce the correlation.
% (see Figure \ref{fig1_ms}). 
The correlation between activity and attractiveness is quantified by the Spearman rank coefficient defined as~\cite{label42}.
\begin{equation}
m_s=1-6\frac{\sum_{i=1}^{V}\triangle_i^2 }{V(V^2-1)},
\label{eq1}
\end{equation}
where $V$ is the network size. $m_s$ is approximately 0 when the nodes' activity and attractiveness are randomly matched, and $m_s$ equals 1 when the nodes' activity and attractiveness are in the same order. In this case, the correlation between activity and attractiveness is the maximum, i.e., the more active a node is, the greater its attractiveness is.
When the node's activity is sorted in ascending order and the attractiveness is sorted in descending order, and then matched one-to-one, than the difference between them is the largest, and $m_s = -1$, In this way, the large the activity of the node is, the smaller the attractiveness is. If the adjustment starts from a fully positively correlated network, $m_s$ decreases with the increase of the fraction $q$ of adjusted nodes. If starting from a fully negatively correlated network, $m_s$ increases with $q$.
%\begin{figure} [ht!]
%\centering
%\includegraphics[width=\linewidth]{ms.jpg}
%\caption{ {\em Regulating the correlation between node activity and attractiveness.} Here, $q$ represents the fraction of reshuffled nodes, while $m_s$ signifies the correlation between the activity and attractiveness of nodes in the network. As the value of $q$ increases, the strength of the positive correlation decreases, while the strength of the negative correlation increases.}
%\label{fig1_ms}
%\end{figure}

\subsection{SIR dynamics}\label{zj}
Here, we use the SIR model to describe the propagation dynamics of disease within a node. A susceptible node is infected by an infected node at rate $\lambda$, while an infected node recovers at rate $\mu$.

Firstly, we focus on the movement of the particles. Based the mean field idea, let the average number of particles on the nodes with activity $a$ and attractiveness $b$ at moment $t$ be $N_{a,b}$, then we have
\begin{equation}
N_{a,b}(t)=\frac{1}{V_{a,b}}\sum_{i|a_i=a,b_i=b}N_i(t),
\label{eq2}
\end{equation} 
where $V_{a,b}$ denotes the number of nodes with activity of $a$ and attractiveness of $b$. Considering a small time period $\triangle t$, the change of $N_{a,b}$ during this time is given by the following equation.
\begin{equation}
\begin{aligned}
d_{t}N_{a,b}(t)=&-ap\triangle t N_{a,b}(t)+\frac{apl\triangle t}{V\left \langle b \right \rangle}\sum_{a^{'},b^{'}}b^{'}N_{a^{'},b^{'}}(t) \\
&-\frac{plb\triangle t\left \langle a \right \rangle N_{a,b}(t)}{\left \langle b \right \rangle }
+\frac{pb\triangle t}{V\left \langle b \right \rangle}\sum_{a^{'},b^{'}}a^{'}N_{a^{'},b^{'}}(t),
\end{aligned}
\label{eq2}
\end{equation} 
where $\left \langle a \right \rangle$ and $\left \langle b \right \rangle $ are the average activity and the average attractiveness of the nodes in the network, respectively. The first term on the right-hand side of Eq. (\ref{eq2}) represents the number of particles released by an active node with activity $a$ and attractiveness $b$. The second term indicates the number of particles attracted by the node with activity $a$ and attractiveness $b$ from other $l$ nodes. The third term represents when the node with activity $a$ and attractiveness is connected to other active nodes and delivers particles to these nodes. The fourth term represents the case when other active nodes connect to and deliver $p\triangle t$ particles to the node with activity $a$ and attractiveness $b$.

When the steady state is reached, $\lim_{t \to \infty} \frac{dN_{a,b}(t)}{dt}=0$, which yields
\begin{equation}
N_{a,b}=\frac{al\sigma +b\chi}{V(\left \langle b \right \rangle a +\left \langle a \right \rangle bl )},
\label{eq3}
\end{equation}
where $\sigma=\sum_{a^{'},b^{'}}b^{'}N_{a^{'},b^{'}}$  and  $\chi=\sum_{a^{'},b^{'}}a^{'}N_{a^{'},b^{'}}$.

Next, we consider the propagation dynamics of the SIR disease. Assume that the basic reproduction number, denoted as $R_0$, is greater than 1 within the population, and only one node is infected at the initial moment. In the initial stage, the number of infected nodes is small, and a tree diagram can be used to
approximate the disease propagation process. Define $D_{a,b}^n$ as the number of infected nodes with an activity of $a$ and an attractiveness of $b$ in the $n$-th generation, which can be determined
from the previous generation $D_{a,b}^{n-1}$. At each time step, there are two ways for a node to become infected by other infected nodes (subpopulations): (1) The node is active and is connected to an infected node. (2) The node is inactive but is connected to an infected node that is active. Considering these two cases, it is obtained that
\begin{equation}
\begin{aligned}
D_{a,b}^n=&\frac{aV_{a,b}}{V\left \langle b \right \rangle} \sum_{a^{'},b^{'}}b^{'}D_{a^{'},b^{'}}^{n-1}(1-R_{0}^{-\beta_{a^{'}b^{'}ab}}) (1-\frac{D_{a,b}^{n-1}}{V_{a,b}} )\\
&+\frac{bV_{a,b}}{V\left \langle b \right \rangle} \sum_{a^{'},b^{'}}a^{'}D_{a^{'},b^{'}}^{n-1}(1-R_{0}^{-\beta_{a^{'}b^{'}ab} }) (1-\frac{D_{a,b}^{n-1}}{V_{a,b}} )
\end{aligned}
\label{eq26}
\end{equation}
where $V_{a,b}$ is the number of nodes with an activity of $a$ and an attractiveness of $b$. $\beta_{a^{'}b^{'}ab}$ is the average number of infected particles that migrate from nodes with activity $a^{'}$ and attractiveness $b^{'}$ to nodes with an activity of $a$ and attractiveness $b$. $R_{0}^{-1}$ denotes the probability the disease vanishes after an infected individual invades a population filled with susceptible individuals. In this case, the probability of an epidemic outbreak at the destination node is $1-R_{0}^{-\beta_{a^{'}b^{'}ab}}$. $1-D_{a,b}^{n-1}/V_{a,b}$ is the probability that a node with an activity $a$ and an attractiveness $b$ is not infected in the $n-1$ generations.

On a static network, the average number of infected particles moving for an infected node with degree $k$ and number of particles $N_k$ to a neighboring node with degree $k^{'}$ is given by $\beta_{k,k^{'}}=Z_{k,k^{'}}\alpha N_k/\mu$, where $Z_{k,k^{'}}$ is the probability a particle from node $k$ to node $k^{'}$, and $\alpha N_k$ is the total number of infected particles generated at $k$ nodes in an epidemic outbreak ($\alpha$ is a parameter related to epidemic model and related parameters). Here, $\alpha \approx \frac{2(R_0-1)}{R_0^2}$~\cite{label20}. Similarly, in the activity-driven network model, there are
\begin{equation}
\beta_{a^{'}b^{'}ab}  =p\alpha N_{a^{'},b^{'}}/\mu.
\label{eq27}
\end{equation}

Considering $l=1$, i.e., the degree of each node is 1. $N_{a^{'},b^{'}}$ is the average number of particles on nodes with activity $a^{'}$ and attractiveness $b^{'}$.

To solve equation (\ref{eq26}), some further approximations are made. Assume that the value of $R_0$ is around 1, so $(1-R_{0}^{-\beta_{a^{'}b^{'}ab}})\sim \beta_{a^{'}b^{'}ab}(R_0-1)$. Since only a few nodes are infected in the initial stage, $(1-\frac{D_{a,b}^{n-1}}{V_{a, b}})\sim 1$. Thus the Eq. (\ref{eq26}) can be rewritten as
\begin{equation}
D_{a,b}^n=\frac{aV_{a,b}}{V\left \langle b \right \rangle}\Omega \sum_{a^{'},b^{'}}b^{'}D_{a^{'},b^{'}}^{n-1}N_{a^{'},b^{'}}+\frac{bV_{a,b}}{V\left \langle b \right \rangle}\Omega \sum_{a^{'},b^{'}}a^{'}D_{a^{'},b^{'}}^{n-1}N_{a^{'},b^{'}},
\label{eq28}
\end{equation} 
where $\Omega=\alpha p (R_0-1)/\mu$. To obtain a closed expression, define
\begin{equation}
\theta^{n}=\sum_{a,b}bD_{a,b}^{n-1}N_{a,b}, \ \xi^{n}=\sum_{a,b}aD_{a,b}^{n-1}N_{a,b},
\label{eq29}
\end{equation}
and
\begin{equation}
\delta=\sum_{a,b}ab\frac{V_{a,b}}{V}N_{a,b},\ \Phi=\sum_{a,b}a^2\frac{V_{a,b}}{V}N_{a,b},\ \kappa=\sum_{a,b}b^2\frac{V_{a,b}}{V}N_{a,b}.
\label{eq30}
\end{equation}

Multiplying both sides of Eq. (\ref{eq28}) by $bN_{a,b}$, and summing $a$ and $b$, we get
\begin{equation}
\theta^{n}=\frac{\delta \Omega}{\left \langle b \right \rangle} \theta^{n-1}+\frac{\kappa \Omega}{\left \langle b \right \rangle} \xi^{n-1}.
\label{eq31}
\end{equation}

Multiplying both sides of Eq. (\ref{eq28}) by $aN_{a,b}$, and summing $a$ and $b$, we get
\begin{equation}
\xi^{n}=\frac{\Phi \Omega}{\left \langle b \right \rangle} \theta^{n-1}+\frac{\delta \Omega}{\left \langle b \right \rangle} \xi^{n-1}.
\label{eq32}
\end{equation}

Under the assumption of continuous time, Eqs. (\ref{eq31}) and (\ref{eq32}) can be rewritten as
\begin{equation}
\partial_n \theta=(\frac{\delta \Omega}{\left \langle b \right \rangle}-1) \theta^{n-1}+\frac{\kappa \Omega}{\left \langle b \right \rangle} \xi^{n-1},
\label{eq33}
\end{equation}
and
\begin{equation}
\partial_n \xi=\frac{\Phi \Omega}{\left \langle b \right \rangle} \theta^{n-1}+(\frac{\delta \Omega}{\left \langle b \right \rangle}-1) \xi^{n-1}.
\label{eq34}
\end{equation}

The Jacobi matrix of the above linear differential equation is
\begin{equation}
J={\left[
\begin{array}{cc}
\frac{\delta \Omega}{\left \langle b \right \rangle}-1 & \frac{\kappa \Omega}{\left \langle b \right \rangle}\\
\frac{\Phi \Omega}{\left \langle b \right \rangle} & \frac{\delta \Omega}{\left \langle b \right \rangle}-1
\end{array}
\right]}
\label{eq35}
\end{equation}
its eigenvalue is
\begin{equation}
\Lambda_{(1,2)}=\frac{\delta \Omega -\left \langle b \right \rangle\pm \Omega\sqrt{\kappa \Phi}}{\left \langle b \right \rangle},
\label{eq36}
\end{equation}
an epidemic outbreak is conditioned on the maximum eigenvalue being greater than 0, thus there is
\begin{equation}
\delta \Omega +\Omega\sqrt{\kappa \Phi} >  \left \langle b \right \rangle,
\label{eq37}
\end{equation}
this gives the critical migration rate of the particle as
\begin{equation}
p^*=\frac{\mu}{\alpha (R_0-1)}\frac{\left \langle b \right \rangle}{\delta +\sqrt{\kappa \Phi}}.
\label{eq37}
\end{equation}

When $p>p^*$, the epidemic can break out in the metapopulation network. Otherwise, the epidemic can only break out on a few nodes. It is worth mentioning that the threshold of the migration rate of particles depends on $\delta$, $\kappa$ and $\Phi$, indicating that it is not only related to the activity and attractiveness of nodes, but also to their correlation.

For the more general case of $l>1$, the critical migration rate of the particle can be solved in a similar way. The only difference is that for the active node, $\beta_{a^{'}b^{'}ab}$ becomes
\begin{equation}
\beta_{a^{'}b^{'}ab}=\frac{p\alpha N_{a^{'},b^{'}}}{l\mu}.
\label{eq40}
\end{equation}

For inactive nodes, $\beta_{a^{'}b^{'}ab}$ takes the same value as in the case of $l=1$. Then Eq. \ref{eq28} can be rewritten as
\begin{equation}
D_{a,b}^n=\frac{alV_{a,b}}{V\left \langle b \right \rangle}\Omega \sum_{a^{'},b^{'}}b^{'}D_{a^{'},b^{'}}^{n-1}N_{a^{'},b^{'}}+\frac{bV_{a,b}}{V\left \langle b \right \rangle}\Omega \sum_{a^{'},b^{'}}a^{'}D_{a^{'},b^{'}}^{n-1}N_{a^{'},b^{'}}.
\label{eq41}
\end{equation}

Similarly, the threshold of migration rate can be obtained as
\begin{equation}
p^*=\frac{\mu}{\alpha (R_0-1)}\frac{2\left \langle b \right \rangle}{(l+1)\delta +\sqrt{4\kappa \Phi l + (l-1)^2\delta^2}}.
\label{eq42}
\end{equation}

Obviously, when $l=1$, Eq. (\ref{eq42}) reduces to the Eq. (\ref{eq28}).

\section{Simulation results}
\subsection{Spreading dynamics in time-varying networks}\label{sec1}
\begin{figure} [ht!]
	\centering
	\includegraphics[width=\linewidth]{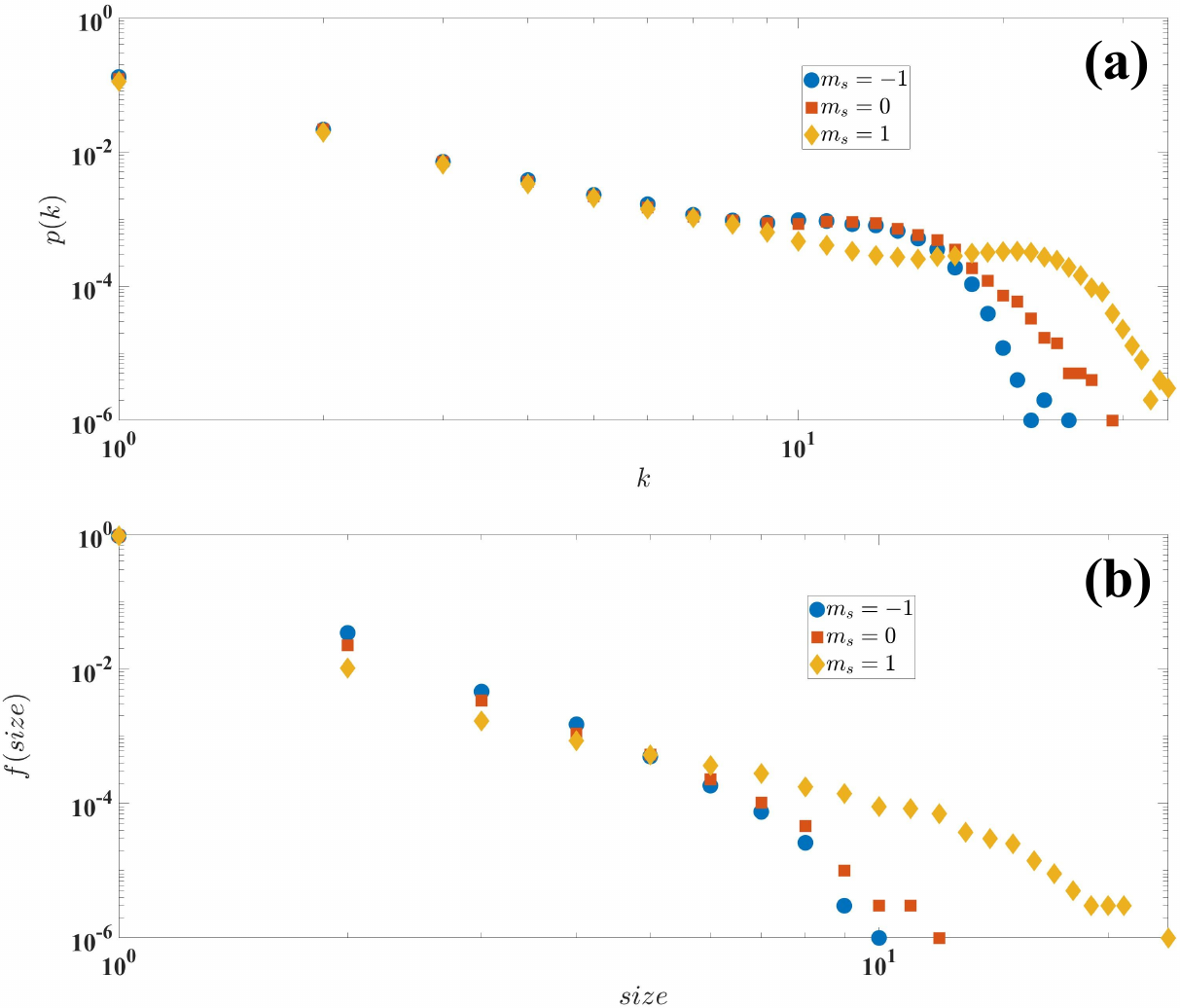}
	\caption{ {\em Topology features of the time-varying population network.} (a) Degree distribution of different correlated networks with number of nodes $V=1000$ in an aggregated network with $T=20$; (b) Subgraph distribution of instantaneous networks with different correlation. Infection rate $\lambda=0.011$, recovery rate $\mu=0.01$, and number of connected edges of each node $l=3$. $N=1000$, $V=1000$. $F(a)\propto a^{-2}$ and $Q(b)\propto b^{-2}$.}
	\label{fig1_du}
\end{figure}

Here, we first present the structural features of the time-varying metapopulation network with activity and attractiveness. Figure \ref{fig1_du}(a) displays the degree distribution $p(k)$ of the aggregating time-varying networks in T=20 times with $V=1000$ nodes. We observe that the maximum degree in the network fully positively correlated ($m_s=1$) network is larger than that of the negatively correlated network and uncorrelated network, while the fraction of small degree nodes is smaller than that of the other two networks.

\begin{figure} [ht!]
	\centering
	\includegraphics[width=\linewidth]{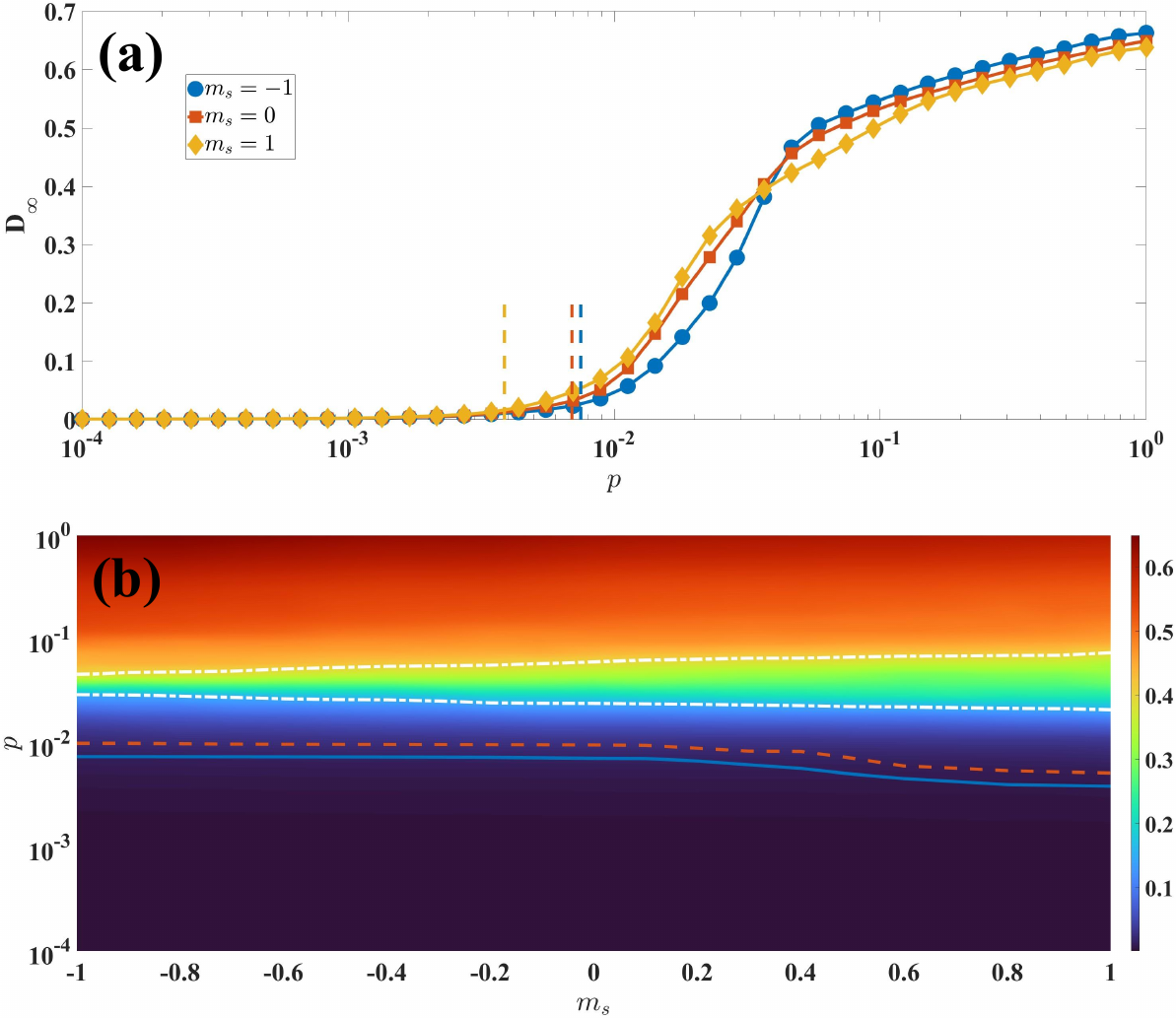}
	\caption{ {\em The impact of the correlation between activity and attractiveness on the spread of SIR-type disease.} (a) Dependence of the fraction of infected subpopulations $D_{\infty}$ on the migration rate $p$ at steady state, where the vertical line represents the theoretical predictions; (b) Dependence of $D_{\infty}$ on $p$ and $m_s$. The dashed line represents the simulated threshold and the solid line represents the theoretical threshold. The white dotted lines represent the different contours.} %Propagation rate $\lambda=0.011$,recovery rate $\mu=0.01$, number of connected edges $l=3$. $N=1000$, $V=1000$. $F(a)\propto a^{-2}$ and $Q(b)\propto b^{-2}$.}
\label{fig1}
\end{figure}

From figure \ref{fig1_du}(b), it can be seen that the size of the largest connected component in instantaneous positively connected network is larger than that of the other two networks, while the fraction of small components in the negatively correlated network is larger than that in the positively correlated network.  This is because in the positively correlated network, nodes with high activity have a large attractiveness. These nodes can connect to and be connected by more nodes, and thus become large degree nodes. Active nodes are more likely to connect to each other and form groups (i.e., large connected
components appear). While in the negatively correlated networks, since active nodes have small attractiveness, the active nodes are not mutually well connected but connect to inactive nodes with large attractiveness. Thus the network is fragmented and have many disconnected small subgraphs, leading a high fraction of small subgraphs in the network. 
Figure \ref{fig1}(a) illustrates the dependence of the infected subpopulation $D_{\infty}$ on the migration rate $p$ in different correlated networks. The critical migration rate of particles, as predicted by Eq. (\ref{eq37}) and Eq. (\ref{eq42}), are represented by the vertical lines. It is evident that the simulation results and analytically predicted results are in good agreement.

Moreover, we observed that both the migration threshold and disease prevalence decrease with the increase in correlation $m_s$ between activity and attractiveness. This is because with the increase of correlation $m_s$, nodes are more likely to form groups, which makes disease spreading more easily. On the other hand, as the correlation increases, the outbreak size $D_{\infty}$ is large when $p$ ($>$ $p_c$) is small, and small when $p$ ($\gg p_c$) is large. This is because as correlation increases, the size of the small group increases. For small values of $p$, disease spreads easily in a large group, and the prevalence $D_{\infty}$ increases. As for large values $p$ ($\gg p_c$), the disease is more likely to spread between large-degree nodes in small groups and to neighbors of large-degree nodes, but hardly infects the large number of small-degree nodes, leading to smaller $D_{\infty}$. Figure \ref{fig1_du} explains this phenomenon to some extent. Figure \ref{fig1}(b) shows the dependence of $D_{\infty}$ on $p$ and $m_s$. The results indicate that the migration threshold decreases as the correlation $m_s$ increases. However, the infection density increases with increasing correlation when $p$ ($>p_c$) is small, and decreases with increasing correlation when $p$ ($\gg$$p_c$) is large. The theoretical migration thresholds agrees with the simulation results in different correlated networks.

\begin{figure} [ht!]
	\centering
	\includegraphics[width=1\linewidth]{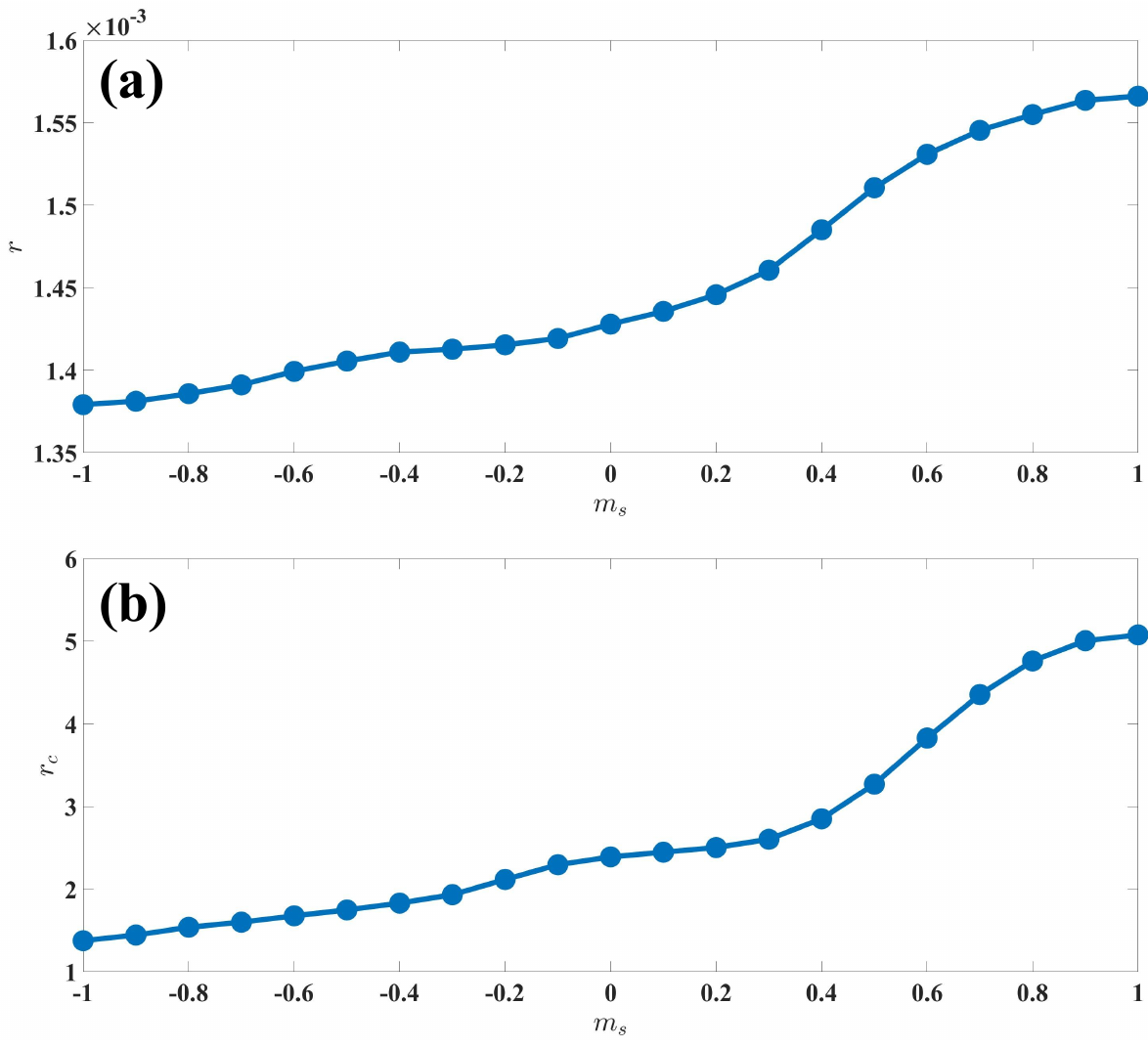}
	\caption{ {\em Variations of the order paramters $ r $ and $ r_c $ on different correlated networks.} (a) Variation of the order parameter $ r $. (b) Variation of the order parameter $ r_c $. Migration rate $ p = $ 0.5.} %, propagation rate $ \lambda = $ 0.011 , recovery rate $ \mu = $ 0.01 , number of connected edges $ l = $ 3 . $ N = $ 1000 , $ V = $ 1000 . $ F(a)\propto a^{-2} $ and $ Q(b)\propto b^{-2} $ .}
\label{fig_4_1_lizi}
\end{figure}

We also study the motion of particles within the network. The distribution of infected particles in the network is described by the order parameter $ r = \sum_{i=1}^{V} \left( \frac{I_i}{\sum_{i=1}^{V} I_i} \right)^2 $, where $I_i$ is the number of infected particles in subpopulation $i$. A larger $ r $ implies a more heterogeneous distribution of infected particles, meaning that a large
number of infected particles exist in a few subpopulations. Furthermore, by multiplying the activity of a node with its attractiveness, i.e., $ K_i = a_i * b_i $, where $a_i$ and $b_i$ represent the activity and attractiveness of node $i$, respectively. This helps in characterizing the importance of nodes in structure and propagation. A larger $ K $ is, the more important
the node is. Take $ z $ nodes with the largest $ K $ value and the total number of infected particles on these $ z $ nodes is defined as $ I_z $. Another parameter $ r_c = \frac{I_z}{z} / \frac{\sum_{i=1}^{V} I_i}{V} $ is introduced to characterise the distribution of infected particles within the central node, which is expressed as the ratio of the average number of infected particles in the central node to the average
number of infected particles in all nodes of the network.

When $ r_c $ much larger than 1, the infected
particles are concentrated in the central node. Figure \ref{fig_4_1_lizi} illustrates the depndence of the ordinal parameters $r$ and $r_c$ on the correlation $m_s$. It can be observed that both $ r $ and $ r_c $ increase with increasing correlation. This
indicates that by increasing the correlation between activity
and attractiveness, the distribution of infected particles will
become more heterogeneous and the infected particles cluster
at the central nodes of the network. This is because
active nodes (with larger $ K $ values) in a positively correlated
network are repeatedly connected to each other and particles
move back and forth between these nodes, which leads to the
concentration of infected particles at these important nodes.

\begin{figure} [ht!]
	\centering
	\includegraphics[width=\linewidth]{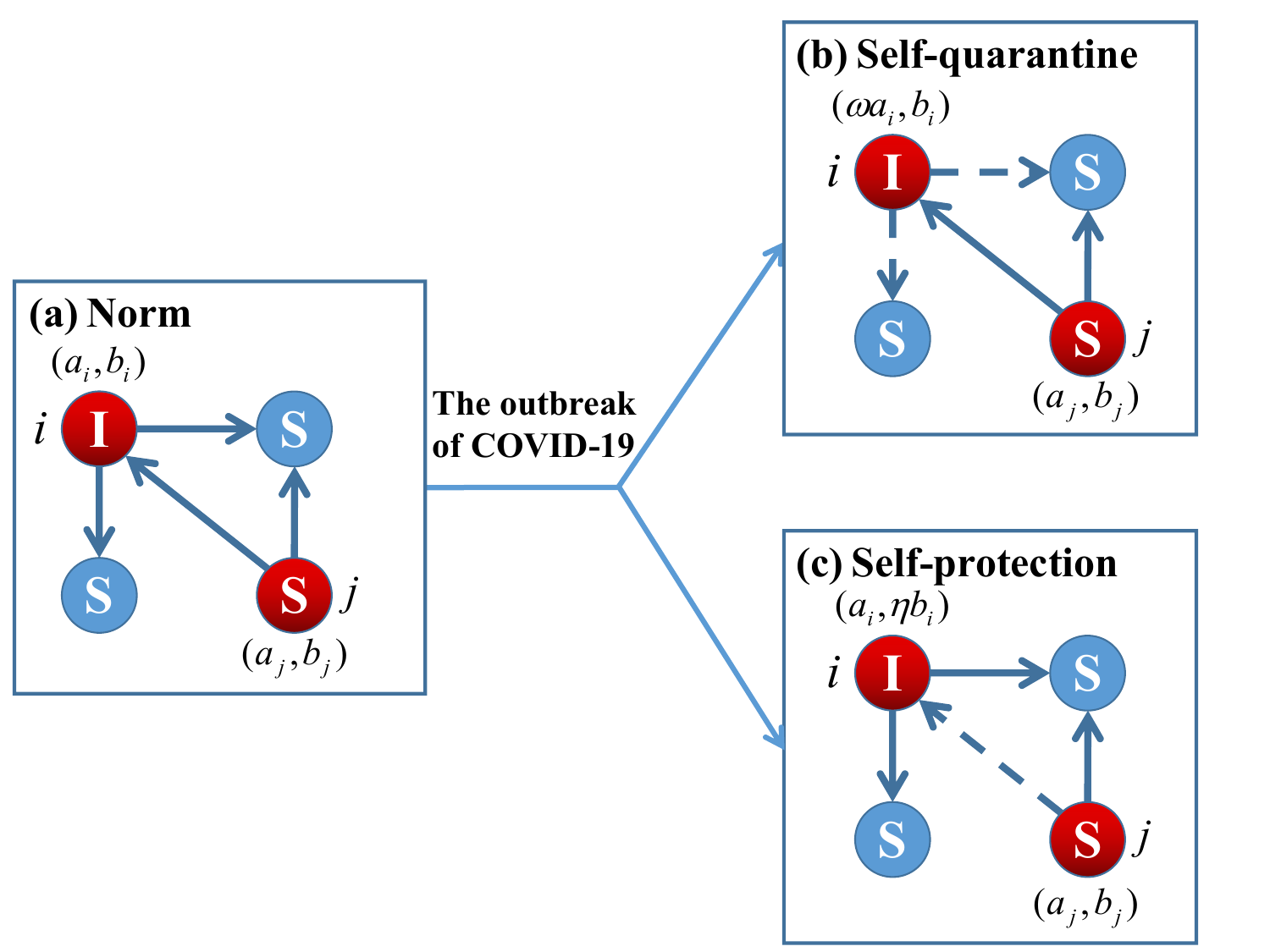}
	\caption{ {\em Schematic diagram illustrating the behavioral changes in populations following the epidemic outbreak.} The red filled circles represent active populations, while the blue filled circles represent inactive populations. The purple solid and dashed lines indicate connections that remain unchanged and connections that are deleted under different behavioral changes, respectively. (a) No behavioral change; (b) Self-isolation of the infected population by reducing their activity; (c) Self-protection of susceptible populations by reducing the attractiveness of the infected population.}
	\label{fig2}
\end{figure}

\subsection{The impact of non-pharmaceutical interventions}
When COVID-19 swept across the globe, countries around the world experienced substantial losses. During the pandemic, it was imperative to implement a series of strict control measures to curb the spread of the virus. Authorities of the city with an outbreak of disease adopt a lockdown strategy to restrict travel of residents, and the city becomes less active. This is called the self-isolation of the infected population [as shown in Figure \ref{fig2}(b)]. Additionally, individuals in cities without an outbreak of epidemic suspend travel to cities experiencing outbreaks to avoid infection. Thus the attractiveness of cities with outbreaks is reduced. We call this behavior as self-protection of susceptible populations [as depicted in Figure \ref{fig2}(c)].

We studied the impacts of behaviors in non-pharmaceutical interventions on epidemics within a time-varying metapopulation networks with attractiveness. Specifically, self-isolation
of infected populations reduced their activity and
the number of links generated, and self-protective behaviors
of susceptible populations reduced the attractiveness of infected
populations. Here, we consider behavioral changes
in epidemic spreading, including self-isolation and self-protection. 
Due to self-isolation of the infected population,
its activity is adjusted by the factor $\eta$ ($\eta a$). In addition, the
attractiveness of the infected population is reduced to $\omega b$
due to self-protection of the susceptible populations. We can
rewrite the propagation dynamics equation as 
\begin{equation}
\begin{aligned}
D_{a,b}^n=&\frac{aV_{a,b}}{V\left \langle b_{actual} \right \rangle} \sum_{a^{'},b^{'}}\omega b^{'}D_{a^{'},b^{'}}^{n-1}(1-R_{0}^{-\beta_{a^{'}b^{'}ab}}) (1-\frac{D_{a,b}^{n-1}}{V_{a,b}} )\\
&+\frac{bV_{a,b}}{V\left \langle b_{actual} \right \rangle} \sum_{a^{'},b^{'}}\eta a^{'}D_{a^{'},b^{'}}^{n-1}(1-R_{0}^{-\beta_{a^{'}b^{'}ab} }) (1-\frac{D_{a,b}^{n-1}}{V_{a,b}} ).
\end{aligned}
\label{eq58}
\end{equation}

Since the infected population is negligible around the threshold, we can assume $\left \langle b_{actual} \right \rangle \approx \left \langle b \right \rangle$. Then similar to the method in section \ref{zj}, we obtain the critical migration rate under NPI as:

when $\eta \ne\omega$,  and
\begin{equation}
p^*=\frac{\mu}{\alpha (R_0-1)}\frac{2\left \langle b \right \rangle}{\delta(\omega+\eta) +\sqrt{(\omega-\eta)^2 \delta^2 + 4\kappa \Phi \eta \omega}}
\label{eq59}
\end{equation}

when $\eta =\omega$,  and
\begin{equation}
p^*=\frac{\mu}{\alpha (R_0-1)}\frac{\left \langle b \right \rangle}{\eta(=\omega)(\delta +\sqrt{\kappa \Phi})}
\label{eq60}
\end{equation}

\subsubsection{Impact of non-pharmaceutical intervention (NPI) policies on different correlated networks}
\begin{figure} [ht!]
	\centering
	\includegraphics[width=\linewidth]{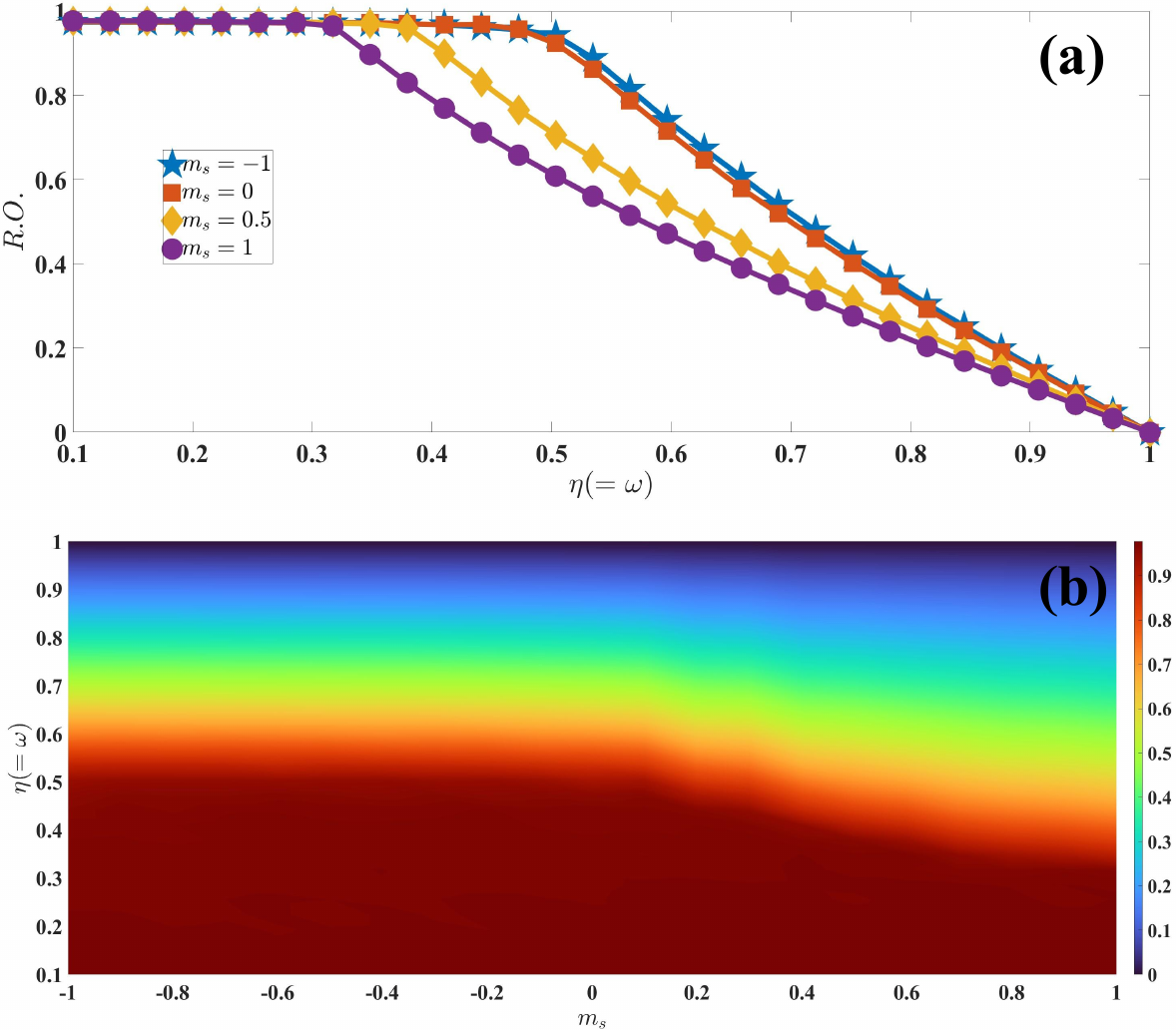}
	\caption{ {\em Impact of non-pharmaceutical intervention (NPI) policies on the spread of disease on different correlated networks.} (a) Effect of NPI on the epidemic in four correlated networks ($\eta$ and $\omega$ decrease by the same amount). $m_s$ is the correlation of node activity and its attractiveness. The horizontal coordinate indicates the intensity of NPI implementation. The lower the value, the higher the intensity. The vertical coordinate indicates the reduction in infection density with NPI policy implemented compared to no NPI implemented. The higher the value, the more effective the suppression of the epidemics. (b) The reduction in the final infection size with the implementation of NPI policy (simultaneous reduction of $\eta$ and $\omega$) on different correlated networks compared to that without the implementation of the NPI policy. The red area indicates the fraction of reduction in the size of the infected population close to 100\%.} %Migration rate $p=1$, transmission rate $\lambda=0.011$,recovery rate $\mu=0.01$. $N=1000$, $V=1000$. $F(a)\propto a^{-2}$ and $Q(b)\propto b^{-2}$. The number of connected edges $l=3$.}
	\label{fig4}
\end{figure}

In section \ref{sec1}, we studied the impact of correlations between the activity and attractiveness of subpopulations on disease transmission. Here, we further investigate the effect of NPI policy on the spread of the epidemic on different correlated networks. Figure \ref{fig4}(a) shows the reduction in the final infection size under NPI policy (meanwhile reducing $\eta$ and $\omega$) on four correlated networks. The reduction ratio is represented as follows:
%$pro=\frac{D_{\infty}^{\eta=\omega=1}-D_{\infty}^{\eta=\omega}}{D_{\infty}^{\eta=\omega=1}} \times \% $. 
\begin{equation}
	R.O.=\frac{D_{\infty}^{\eta=\omega=1}-D_{\infty}^{\eta=\omega}}{D_{\infty}^{\eta=\omega=1}} \times 100\%. 
	\label{eq61}
\end{equation}

We observed that when $\eta=\omega=0.5$, the reduction ratio on the negatively correlated network is close to 100\%, whereas on the positively correlated network, it is only 60\%. This suggests that implementing the NPI policy on negatively correlated network is more effective and can significantly reduce the outbreak size. This is because when the activity of subpopulation networks is positively correlated with attractiveness, well-connected groups exist within the network. In contrast, on negatively correlated networks, the subpopulation network is smaller and the network structure is sparse. Consequently, when implementing the NPI policy, similar to network disintegration (removal of nodes and edges), the connectivity of the negatively correlated network is disrupted more than that of the positively correlated network, which is more robust. Therefore, implementing the NPI policy results in a more fragmented negatively correlated network, thus is more effective in suppressing the spread of the epidemic.
Figure \ref{fig4}(b) displays the reduction in the final infection size under the NPI policy (meanwhile reducing $\eta$ and $\omega$) on different correlated networks. The red area indicates a nearly 100\% reduction in the size of the infected population, implying that the strict NPI policy ($\eta$ and $\omega$ are close to 0) can effectively suppress the epidemic.

\begin{figure} [ht!]
	\centering
	\includegraphics[width=\linewidth]{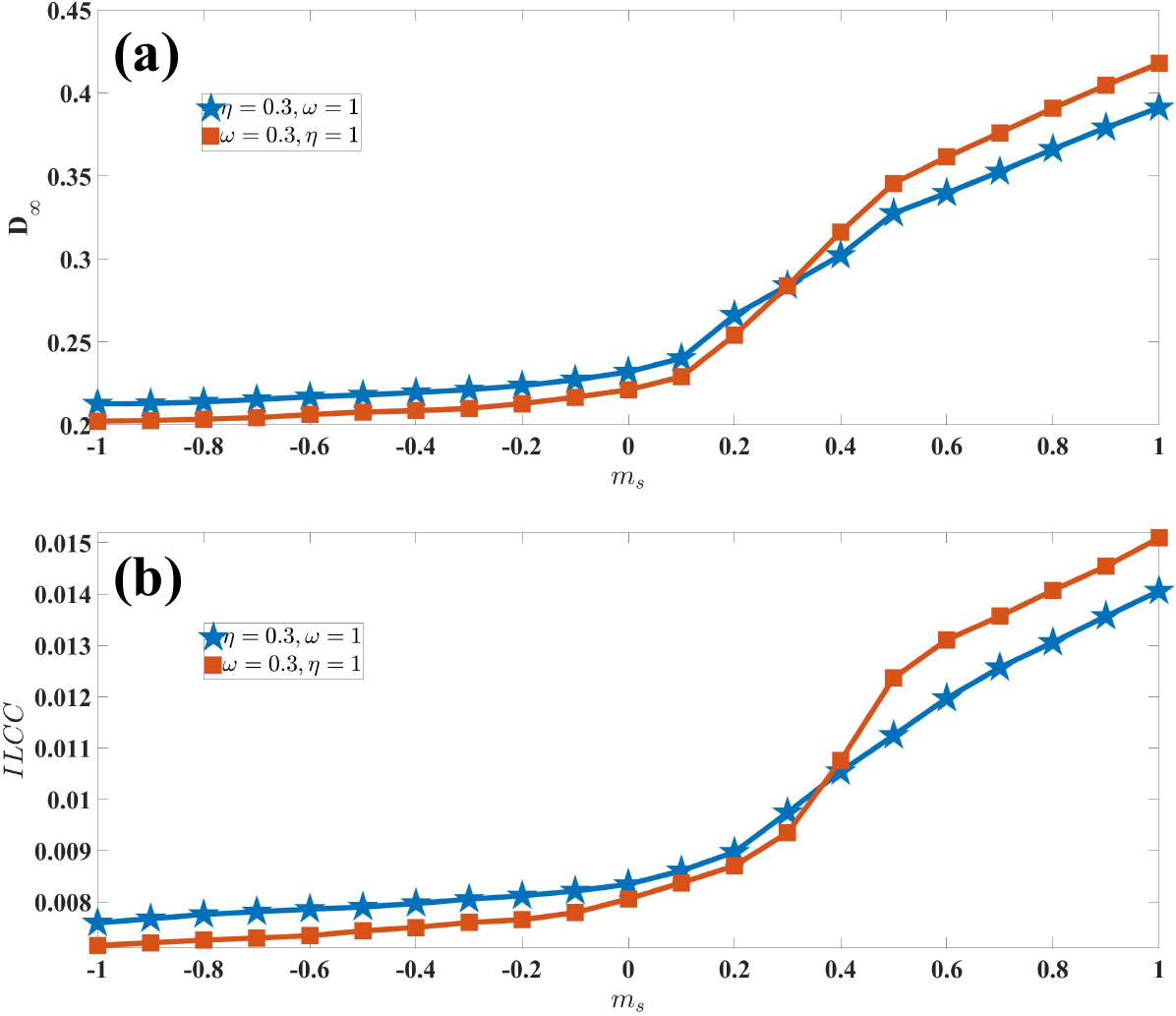}
	\caption{ {\em The effect of a single non-pharmaceutical intervention policy on the spread of an epidemic on different networks of association.} (a) The effect of self-isolation or self-protection on the spread of outbreaks in networks with different correlations. (b) Average maximum connected subgraph size for instantaneous networks with different correlations.} %Migration rate $p=1$, propagation rate $\lambda=0.011$,recovery rate $\mu=0.01$. $N=1000$, $V=1000$. $F(a)\propto a^{-2}$ and $Q(b)\propto b^{-2}$. The number of connected edges $l=3$.}
	\label{fig5}
\end{figure}

In addition, we investigated the effect of single non-pharmaceutical intervention policy on the spread of the epidemic on different correlated networks. As shown in Figure \ref{fig5}(a), the stronger the positive correlation is, the better the self-protection of the susceptible population (reducing $\eta$) inhibits the spread of the epidemic than that of the self-isolation of the infected population (reducing $\omega$). In contrast, when correlations tend to be negative, self-isolation strategies for infected populations are more effective compared to self-protection strategies for susceptible populations. Figure \ref{fig5}(b) shows the average size of the largest connected subgraph of the instantaneous networks with different correlations after implementing a single non-pharmaceutical intervention policy. We find that the size of the largest connected component is larger under self-protection policy than under self-isolation when the correlation is negative. With the increase of correlation, the size of the largest connected component under self-protection policy becomes smaller than that under the self-isolation policy.

This is because when the correlation between the activity of a subpopulation and its attractiveness is positive, the more active nodes are, the more attractive they are. If the most active population $i$ (and also the most attractive) becomes the infected population and it adopt a self-isolation strategy (reducing $\omega$), then its activity decreases, while attractiveness remains. This infected population $i$ will not actively connect to other nodes, but other active nodes will still choose to connect to it because of its high attractiveness. If the infected population $i$ becomes less attractive (reducing $\eta$), while keeps its activity, it will actively connect to other nodes but other active nodes do not choose to connect to it. In general, the node with high attractiveness has a higher degree than the node with high activity. This indicates that the infected population $i$ decreasing $\omega$ has a larger number of edges than that decreasing $\eta$. So the infected population decreasing $\omega$ is more likely to infect other susceptible populations than the infected population decreasing $\eta$. Therefore, in positively correlated networks, self-protection is more effective in suppressing epidemics than self-isolation of infected populations.

When the activity of a subpopulation is negatively correlated with its attractiveness, active nodes have less attractiveness. If the most active population $i$ (its attractiveness is minimal) becomes infected, it only adopts a self-isolation strategy to control the spread of the epidemic. With the decrease of its activity, it does not actively connect with other populations, and other populations also do not connect with it (due to low attractiveness). If the infected population $i$ decreases $\eta$, its attractiveness decreases (its own attractiveness is small) and its activity remains unchanged, when it will actively connect to other populations. Thus, on a negatively correlated network, infected populations $i$ have a greater number of edges reducing $\eta$ than that of reducing $\omega$, and therefore is more likely to infect other populations. This implies that when the correlation of activity and attractiveness is negative, isolating infected populations is more effective in suppressing epidemics than protecting susceptible populations.

\subsubsection{Impact of self-isolation or self-protection on epidemics}
\begin{figure*} [ht!]
\centering
\includegraphics[width=\linewidth]{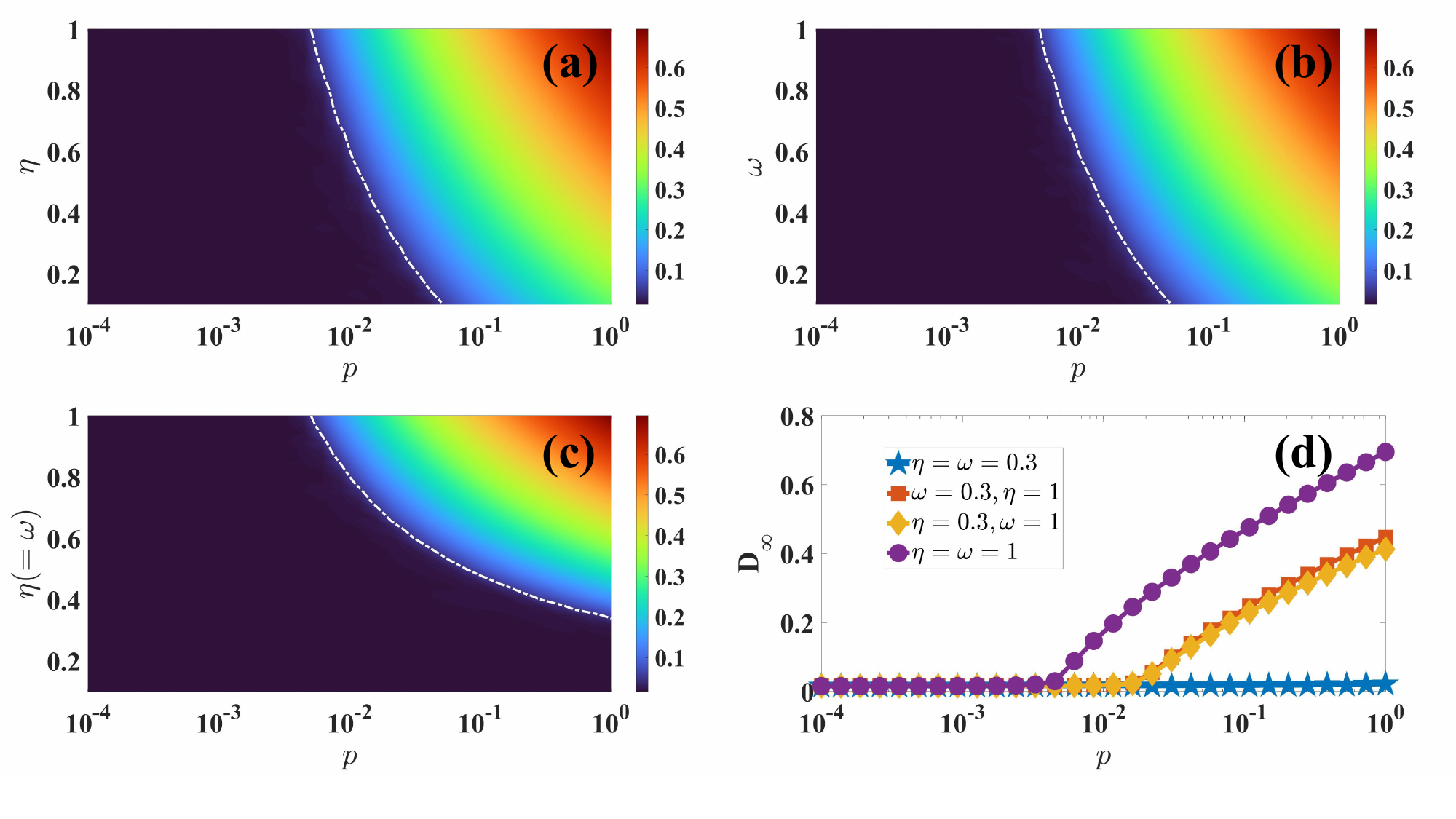}
\caption{ {\em The impact of self-isolation or self-protection on epidemics in a perfectly positively correlated network with $m_s=1$.} (a) Self-protection, infected population densities $D_{\infty}$ versus $\eta$ and $p$, $\omega=1$. (b) Self-isolation, infected population density $D_{\infty}$ versus $\omega$ and $p$, where $\eta=1$. (c) Effect of joint changes in self-isolation ($\eta$) and self-protection ($\omega$) on epidemic dynamics. (d) Density of infected populations $D_{\infty}$ versus $p$ under various behavioral modifications.} %Transmission rate$\lambda=0.011$,recovery rate$\mu=0.01$. $N=1000$, $V=1000$. $F(a)\propto a^{-2}$ and $Q(b)\propto b^{-2}$. The number of connected edges $l=3$.}
\label{fig6}
\end{figure*}

We validated the theoretical predictions from Eq. (\ref{eq59}) and Eq. (\ref{eq60}) through numerical simulations. Additionally, we further investigated the impact of non-pharmaceutical intervention policies on epidemic dynamics in positively correlated network ($m_s=1$). Figures \ref{fig6}(a)-(b) display the impact of $\eta$ or $\omega$ on the epidemic. It can be seen that the migration thresholds derived from Eq. (\ref{eq59}) is consistent with the Monte Carlo simulation results. As seen in Figure \ref{fig6}(a), the migration threshold $p_c$ increases as $\eta$ decreases, indicating that self-isolation can delay the onset of the epidemic. This same effect of self-protection is observed in Figure \ref{fig6}(b). Furthermore, interchanging the self-isolation factor $\omega$ and the self-protection factor $\eta$ results in the same threshold expression. Thus the same pair of parameter values for $\omega$ and $\eta$ yields the same migration threshold. This suggests that self-isolation of infected populations and self-protection of susceptible populations can delay the outbreak to the same extent when implemented with the same intensity.

Figure \ref{fig6}(c) illustrates the impact of joint changes in self-isolation and self-protection on migration threshold and infection size, where both $\eta$ and $\omega$ decrease by the same amount. The migration threshold predicted by our model is very close to the Monte Carlo simulations. In Figure \ref{fig6}(c), as $\eta$ and $\omega$ decrease, the migration threshold increases, and the infection size decreases. This implies that joint changes in self-isolation and self-protection can slow down the spread
of the epidemic. Moreover, the single behavioral change does not result as large in an increase in migration threshold as the combined changes in $\eta$ and $\omega$. This suggests
that joint changes in behaviors further delay the epidemic.
Figure \ref{fig6}(d) displays the dependence of the infected population size $D_{\infty}$ on the migration rate $p$ , the self-isolation factor $\eta$ and the self-protection factor $\omega$. We take the SIR spreading processes without self-isolation and self-protection ($\eta=\omega=1$) as the benchmark. It can be seen that as the values of $\eta$ or $\omega$ decreases, the migration threshold increases, and the density of infected populations decreases. This suggests that both self-isolation and self-protection in reducing the size of infection and delaying the epidemic are effective. Simultaneously adopting self-isolation and self-protection further increase the migration threshold and reduce the density of infected populations.

Furthermore, in the positively correlated network, the same intensity of self-protection of susceptible populations and self-isolation of infected populations leads to the same migration threshold, while the density of infected populations is slightly higher for self-isolation. This is consistent with the results shown in Figure \ref{fig5}.
We can conclude that both the self-isolation of infected populations and the self-protection of susceptible populations increase the migration threshold, thereby delaying the outbreak. However, when it comes to reducing the density of infected populations, their effectiveness varies across different correlated networks. On the positively correlated network, self-protection is more effective, while on the negatively correlated network, self-isolation is more effective.

%\begin{figure*} [ht!]
%\centering
%\includegraphics[scale=0.45]{tupian13.pdf}
%\caption{Effects of different initial conditions on rumor in activity-driven networks.(a-c)The relationship between $\lambda$and $R_{\infty}$under different initial conditions on $m_s=1,0,-1$ network are shown. }
%\label{fig10}
%\end{figure*}

\section{Conclusion}

Due to convenience in transportation, migration of individuals between regions, cities or countries can impact spatial transmission of epidemics. In addition, their migraion paths vary over time. It is necessary to consider the impact of time-varying migration paths on epidemic transmission.
In this paper, we analyze the reaction-diffusion processes in metapopulation networks characterized by a time-varying coupling pattern within an attractive activity-driven framework. We use the SIR model and analytically derive epidemic threshold in metapopulation networks, which agrees well with simulation results. We find that the migration threshold decreased with increasing correlation ($m_s$) between the activity and attractiveness of a node. As the correlation increased, the outbreak size ($D_{\infty}$) increases when $p$ ($>p_c$) was small, and the outbreak size ($D_{\infty}$) decreases when $p$ ($\gg$$p_c$) is large. The distribution of infected particles is heterogeneity on positively correlated networks and is homogeneous on negatively correlated networks.

Then, we associate the non-pharmaceutical intervention behaviors (self-isolation and self-protection) with attributes (activity and attractiveness) of populations. The effects of non-pharmaceutical intervention behaviors on epidemics are studied in time-varying metapopulation model with attractiveness. Specifically, self-isolation of infected populations reduces their activity, while self-protection of susceptible populations led to a decrease in the attractiveness of infected populations. The non-pharmaceutical interventions (self-protection and self-isolation) are more effective in reducing the size of outbreaks on negatively correlated networks than on positively correlated networks. Both self-isolation and self-protection can increase migration threshold to delaying the disease outbreak. However, the effectiveness of a single strategy in reducing the density of infected populations varies in different correlated networks. Self-protection is more effective in positively correlated networks, while self-isolation is more effective in negatively correlated networks. This study investigate the impact of correlations among node attributes on the dynamics on top of it. It aims to enhance our comprehension of the relationship between the characteristics of time-varying networks and population behavior during epidemics, and provides insights for developing effective intervention strategies to control epidemics.

However, this work has several limitations. First, we simply considered the case where the time scale of network evolution was comparable to the time scale of dynamical processes on it and the time scale of individual migration. However, it often takes time for individuals to move across the metapopulation networks, such as traveling from one region to another. We can further study the effects of inconsistent time scales and interactions between individuals during their journeys on epidemic spreading. Second, although we focused on some fundamental aspects of the real world metapopulations, such as the correlation between population activity and attractiveness, as well as behavioral changes, other significant factors is neglected, such as the age distribution within populations. Therefore, it is necessary to consider the intrapopulation demographic characteristics on the spread of diseases throughout the metapopulation network.

\section{Acknowledgments} \label{sec:intro}
This work was supported by the National “ten thousand talents plan” youth top talent project, the National Natural Science Foundation of China (Grant Nos. 12231012, 11975099, 61802321), and the Science and Technology Commission of Shanghai Municipality (Grant No. 14DZ2260800), and
the National Research Foundation of Korea (NRF) 
grant funded by the Korean government (MSIT) (No.~NRF-2022R1A5A1033624 \& 2022R1A2C3011711).

%\bibliography{COVID_19}
\bibliographystyle{naturemag}
%\bibliography{IEEEexample}
\bibliography{time}

\begin{thebibliography}{10}
\expandafter\ifx\csname url\endcsname\relax
  \def\url#1{\texttt{#1}}\fi
\expandafter\ifx\csname urlprefix\endcsname\relax\def\urlprefix{URL }\fi
\providecommand{\bibinfo}[2]{#2}
\providecommand{\eprint}[2][]{\url{#2}}

\bibitem{label32}
\bibinfo{author}{Zhai, Z.-M.}, \bibinfo{author}{Long, Y.-S.},
  \bibinfo{author}{Tang, M.}, \bibinfo{author}{Liu, Z.} \&
  \bibinfo{author}{Lai, Y.-C.}
\newblock \bibinfo{title}{Optimal inference of the start of covid-19}.
\newblock \emph{\bibinfo{journal}{Physical Review Research}}
  \textbf{\bibinfo{volume}{3}}, \bibinfo{pages}{013155} (\bibinfo{year}{2021}).

\bibitem{label38}
\bibinfo{author}{Chinazzi, M.} \emph{et~al.}
\newblock \bibinfo{title}{The effect of travel restrictions on the spread of
  the 2019 novel coronavirus (covid-19) outbreak}.
\newblock \emph{\bibinfo{journal}{Science}} \textbf{\bibinfo{volume}{368}},
  \bibinfo{pages}{395--400} (\bibinfo{year}{2020}).

\bibitem{label39}
\bibinfo{author}{Wu, F.} \emph{et~al.}
\newblock \bibinfo{title}{A new coronavirus associated with human respiratory
  disease in china}.
\newblock \emph{\bibinfo{journal}{Nature}} \textbf{\bibinfo{volume}{579}},
  \bibinfo{pages}{265--269} (\bibinfo{year}{2020}).

\bibitem{label20}
\bibinfo{author}{Colizza, V.} \& \bibinfo{author}{Vespignani, A.}
\newblock \bibinfo{title}{Invasion threshold in heterogeneous metapopulation
  networks}.
\newblock \emph{\bibinfo{journal}{Physical review letters}}
  \textbf{\bibinfo{volume}{99}}, \bibinfo{pages}{148701}
  (\bibinfo{year}{2007}).

\bibitem{label21}
\bibinfo{author}{Colizza, V.}, \bibinfo{author}{Pastor-Satorras, R.} \&
  \bibinfo{author}{Vespignani, A.}
\newblock \bibinfo{title}{Reaction--diffusion processes and metapopulation
  models in heterogeneous networks}.
\newblock \emph{\bibinfo{journal}{Nature Physics}}
  \textbf{\bibinfo{volume}{3}}, \bibinfo{pages}{276--282}
  (\bibinfo{year}{2007}).

\bibitem{label22}
\bibinfo{author}{Meloni, S.} \emph{et~al.}
\newblock \bibinfo{title}{Modeling human mobility responses to the large-scale
  spreading of infectious diseases}.
\newblock \emph{\bibinfo{journal}{Scientific reports}}
  \textbf{\bibinfo{volume}{1}}, \bibinfo{pages}{1--7} (\bibinfo{year}{2011}).

\bibitem{label23}
\bibinfo{author}{dos Reis, E.~F.} \& \bibinfo{author}{Masuda, N.}
\newblock \bibinfo{title}{Metapopulation models imply non-poissonian statistics
  of interevent times}.
\newblock \emph{\bibinfo{journal}{Physical Review Research}}
  \textbf{\bibinfo{volume}{4}}, \bibinfo{pages}{013050} (\bibinfo{year}{2022}).

\bibitem{label24}
\bibinfo{author}{Ye, Y.} \emph{et~al.}
\newblock \bibinfo{title}{Equitable access to covid-19 vaccines makes a
  life-saving difference to all countries}.
\newblock \emph{\bibinfo{journal}{Nature human behaviour}}
  \textbf{\bibinfo{volume}{6}}, \bibinfo{pages}{207--216}
  (\bibinfo{year}{2022}).

\bibitem{label25}
\bibinfo{author}{Tennant, W.~S.} \emph{et~al.}
\newblock \bibinfo{title}{Modelling the persistence and control of rift valley
  fever virus in a spatially heterogeneous landscape}.
\newblock \emph{\bibinfo{journal}{Nature communications}}
  \textbf{\bibinfo{volume}{12}}, \bibinfo{pages}{1--13} (\bibinfo{year}{2021}).

\bibitem{label26}
\bibinfo{author}{Davis, J.~T.} \emph{et~al.}
\newblock \bibinfo{title}{Cryptic transmission of sars-cov-2 and the first
  covid-19 wave}.
\newblock \emph{\bibinfo{journal}{Nature}} \textbf{\bibinfo{volume}{600}},
  \bibinfo{pages}{127--132} (\bibinfo{year}{2021}).

\bibitem{label27}
\bibinfo{author}{Balcan, D.} \emph{et~al.}
\newblock \bibinfo{title}{Multiscale mobility networks and the spatial
  spreading of infectious diseases}.
\newblock \emph{\bibinfo{journal}{Proceedings of the National Academy of
  Sciences}} \textbf{\bibinfo{volume}{106}}, \bibinfo{pages}{21484--21489}
  (\bibinfo{year}{2009}).

\bibitem{label31}
\bibinfo{author}{Kermack, W.~O.} \& \bibinfo{author}{McKendrick, A.~G.}
\newblock \bibinfo{title}{A contribution to the mathematical theory of
  epidemics}.
\newblock \emph{\bibinfo{journal}{Proceedings of the royal society of london}}
  \textbf{\bibinfo{volume}{115}}, \bibinfo{pages}{700--721}
  (\bibinfo{year}{1927}).

\bibitem{label40}
\bibinfo{author}{Pastor-Satorras, R.} \& \bibinfo{author}{Vespignani, A.}
\newblock \bibinfo{title}{Epidemic spreading in scale-free networks}.
\newblock \emph{\bibinfo{journal}{Physical review letters}}
  \textbf{\bibinfo{volume}{86}}, \bibinfo{pages}{3200} (\bibinfo{year}{2001}).

\bibitem{label41}
\bibinfo{author}{Eguiluz, V.~M.} \& \bibinfo{author}{Klemm, K.}
\newblock \bibinfo{title}{Epidemic threshold in structured scale-free
  networks}.
\newblock \emph{\bibinfo{journal}{Physical Review Letters}}
  \textbf{\bibinfo{volume}{89}}, \bibinfo{pages}{108701}
  (\bibinfo{year}{2002}).

\bibitem{label33}
\bibinfo{author}{Barrat, A.} \emph{et~al.}
\newblock \bibinfo{title}{Empirical temporal networks of face-to-face human
  interactions}.
\newblock \emph{\bibinfo{journal}{The European Physical Journal Special
  Topics}} \textbf{\bibinfo{volume}{222}}, \bibinfo{pages}{1295--1309}
  (\bibinfo{year}{2013}).

\bibitem{label34}
\bibinfo{author}{Newman, M. E.~J.}
\newblock \bibinfo{title}{The structure of scientific collaboration networks}.
\newblock \emph{\bibinfo{journal}{Proceedings of the National Academy of
  Sciences}} \textbf{\bibinfo{volume}{98}}, \bibinfo{pages}{404--409}
  (\bibinfo{year}{2001}).
\newblock \urlprefix\url{https://www.pnas.org/content/98/2/404}.
\newblock \eprint{https://www.pnas.org/content/98/2/404.full.pdf}.

\bibitem{label1}
\bibinfo{author}{Holme, P.} \& \bibinfo{author}{Saram{\"a}ki, J.}
\newblock \bibinfo{title}{Temporal networks}.
\newblock \emph{\bibinfo{journal}{Physics reports}}
  \textbf{\bibinfo{volume}{519}}, \bibinfo{pages}{97--125}
  (\bibinfo{year}{2012}).

\bibitem{label2}
\bibinfo{author}{Salath{\'e}, M.} \emph{et~al.}
\newblock \bibinfo{title}{A high-resolution human contact network for
  infectious disease transmission}.
\newblock \emph{\bibinfo{journal}{Proceedings of the National Academy of
  Sciences}} \textbf{\bibinfo{volume}{107}}, \bibinfo{pages}{22020--22025}
  (\bibinfo{year}{2010}).

\bibitem{label3}
\bibinfo{author}{Li, A.}, \bibinfo{author}{Cornelius, S.~P.},
  \bibinfo{author}{Liu, Y.-Y.}, \bibinfo{author}{Wang, L.} \&
  \bibinfo{author}{Barab{\'a}si, A.-L.}
\newblock \bibinfo{title}{The fundamental advantages of temporal networks}.
\newblock \emph{\bibinfo{journal}{Science}} \textbf{\bibinfo{volume}{358}},
  \bibinfo{pages}{1042--1046} (\bibinfo{year}{2017}).

\bibitem{label4}
\bibinfo{author}{Zhang, Y.} \& \bibinfo{author}{Strogatz, S.~H.}
\newblock \bibinfo{title}{Designing temporal networks that synchronize under
  resource constraints}.
\newblock \emph{\bibinfo{journal}{Nature communications}}
  \textbf{\bibinfo{volume}{12}}, \bibinfo{pages}{1--8} (\bibinfo{year}{2021}).

\bibitem{label5}
\bibinfo{author}{Kobayashi, T.}, \bibinfo{author}{Takaguchi, T.} \&
  \bibinfo{author}{Barrat, A.}
\newblock \bibinfo{title}{The structured backbone of temporal social ties}.
\newblock \emph{\bibinfo{journal}{Nature communications}}
  \textbf{\bibinfo{volume}{10}}, \bibinfo{pages}{1--11} (\bibinfo{year}{2019}).

\bibitem{label6}
\bibinfo{author}{Valdano, E.}, \bibinfo{author}{Ferreri, L.},
  \bibinfo{author}{Poletto, C.} \& \bibinfo{author}{Colizza, V.}
\newblock \bibinfo{title}{Analytical computation of the epidemic threshold on
  temporal networks}.
\newblock \emph{\bibinfo{journal}{Physical Review X}}
  \textbf{\bibinfo{volume}{5}}, \bibinfo{pages}{021005} (\bibinfo{year}{2015}).

\bibitem{label7}
\bibinfo{author}{Li, A.} \emph{et~al.}
\newblock \bibinfo{title}{Evolution of cooperation on temporal networks}.
\newblock \emph{\bibinfo{journal}{Nature communications}}
  \textbf{\bibinfo{volume}{11}}, \bibinfo{pages}{1--9} (\bibinfo{year}{2020}).

\bibitem{label9}
\bibinfo{author}{Williams, O.~E.}, \bibinfo{author}{Lacasa, L.},
  \bibinfo{author}{Mill{\'a}n, A.~P.} \& \bibinfo{author}{Latora, V.}
\newblock \bibinfo{title}{The shape of memory in temporal networks}.
\newblock \emph{\bibinfo{journal}{Nature Communications}}
  \textbf{\bibinfo{volume}{13}}, \bibinfo{pages}{1--8} (\bibinfo{year}{2022}).

\bibitem{label10}
\bibinfo{author}{Unicomb, S.}, \bibinfo{author}{I{\~n}iguez, G.},
  \bibinfo{author}{Gleeson, J.~P.} \& \bibinfo{author}{Karsai, M.}
\newblock \bibinfo{title}{Dynamics of cascades on burstiness-controlled
  temporal networks}.
\newblock \emph{\bibinfo{journal}{Nature communications}}
  \textbf{\bibinfo{volume}{12}}, \bibinfo{pages}{1--10} (\bibinfo{year}{2021}).

\bibitem{label17}
\bibinfo{author}{Perra, N.}, \bibinfo{author}{Gon{\c{c}}alves, B.},
  \bibinfo{author}{Pastor-Satorras, R.} \& \bibinfo{author}{Vespignani, A.}
\newblock \bibinfo{title}{Activity driven modeling of time varying networks}.
\newblock \emph{\bibinfo{journal}{Scientific reports}}
  \textbf{\bibinfo{volume}{2}}, \bibinfo{pages}{1--7} (\bibinfo{year}{2012}).

\bibitem{label18}
\bibinfo{author}{Perra, N.} \emph{et~al.}
\newblock \bibinfo{title}{Random walks and search in time-varying networks}.
\newblock \emph{\bibinfo{journal}{Physical review letters}}
  \textbf{\bibinfo{volume}{109}}, \bibinfo{pages}{238701}
  (\bibinfo{year}{2012}).

\bibitem{label35}
\bibinfo{author}{Liu, S.}, \bibinfo{author}{Perra, N.},
  \bibinfo{author}{Karsai, M.} \& \bibinfo{author}{Vespignani, A.}
\newblock \bibinfo{title}{Controlling contagion processes in activity driven
  networks}.
\newblock \emph{\bibinfo{journal}{Physical Review Letters}}
  \textbf{\bibinfo{volume}{112}}, \bibinfo{pages}{118702}
  (\bibinfo{year}{2014}).

\bibitem{label36}
\bibinfo{author}{Perra, N.} \emph{et~al.}
\newblock \bibinfo{title}{Random walks and search in time-varying networks}.
\newblock \emph{\bibinfo{journal}{Physical Review Letters}}
  \textbf{\bibinfo{volume}{109}}, \bibinfo{pages}{238701}
  (\bibinfo{year}{2012}).

\bibitem{label37}
\bibinfo{author}{Moinet, A.}, \bibinfo{author}{Starnini, M.} \&
  \bibinfo{author}{Pastor-Satorras, R.}
\newblock \bibinfo{title}{Burstiness and aging in social temporal networks}.
\newblock \emph{\bibinfo{journal}{Physical Review Letters}}
  \textbf{\bibinfo{volume}{114}}, \bibinfo{pages}{108701}
  (\bibinfo{year}{2015}).

\bibitem{label12}
\bibinfo{author}{Wang, B.}, \bibinfo{author}{Ding, X.} \& \bibinfo{author}{Han,
  Y.}
\newblock \bibinfo{title}{Phase transition in the majority-vote model on
  time-varying networks}.
\newblock \emph{\bibinfo{journal}{Physical Review E}}
  \textbf{\bibinfo{volume}{105}}, \bibinfo{pages}{014310}
  (\bibinfo{year}{2022}).

\bibitem{label30}
\bibinfo{author}{Changruenngam, S.}, \bibinfo{author}{Bicout, D.~J.} \&
  \bibinfo{author}{Modchang, C.}
\newblock \bibinfo{title}{How the individual human mobility spatio-temporally
  shapes the disease transmission dynamics}.
\newblock \emph{\bibinfo{journal}{Scientific Reports}}
  \textbf{\bibinfo{volume}{10}}, \bibinfo{pages}{1--13} (\bibinfo{year}{2020}).

\bibitem{label42}
\bibinfo{author}{Lee, K.-M.}, \bibinfo{author}{Kim, J.~Y.},
  \bibinfo{author}{Cho, W.-k.}, \bibinfo{author}{Goh, K.-I.} \&
  \bibinfo{author}{Kim, I.}
\newblock \bibinfo{title}{Correlated multiplexity and connectivity of multiplex
  random networks}.
\newblock \emph{\bibinfo{journal}{New Journal of Physics}}
  \textbf{\bibinfo{volume}{14}}, \bibinfo{pages}{033027}
  (\bibinfo{year}{2012}).

\end{thebibliography}
\end{document}